\documentclass[%
 reprint,
superscriptaddress,
aps,
prb,
 amsmath,amssymb,
]{revtex4-2}

\setcitestyle{super}
\usepackage{graphicx}% Include figure files
\usepackage{dcolumn}% Align table columns on decimal point
\usepackage{bm}% bold math
\usepackage{booktabs,xcolor,colortbl}
\usepackage{xparse}
\usepackage[toc,page]{appendix}
\usepackage{caption}
\usepackage{subcaption}
\usepackage{float}
\usepackage{afterpage}
\usepackage[version=4]{mhchem}
\usepackage{multirow}
\usepackage{tabularx}

\usepackage{palatino}
\usepackage{todonotes}
\usepackage{color}
\usepackage[normalem]{ulem}

\begin{document}

\title{Discovering Ferroelectric Plastic (Ionic) Crystals in the Cambridge Structural Database: Database Mining and Computational Assessment}

\author{Elin Dypvik Sødahl}
\email{elin.dypvik.sodahl@nmbu.no}
\affiliation{Department of Mechanical Engineering and Technology Management, Norwegian University of
Life Sciences, 1432 Ås, Norway.}

\author{Seyedmojtaba Seyedraoufi}
\affiliation{Department of Mechanical Engineering and Technology Management, Norwegian University of
Life Sciences, 1432 Ås, Norway.}

\author{Carl Henrik Görbitz}
\affiliation{Department of Chemistry, University of Oslo, 0371 Oslo, Norway.}

\author{Kristian Berland}
\email{kristian.berland@nmbu.no}
\affiliation{Department of Mechanical Engineering and Technology Management, Norwegian University of
Life Sciences, 1432 Ås, Norway.}

\date{\today}

\begin{abstract}
Hybrid or organic plastic crystals have the potential as lead-free alternatives to conventional inorganic ferroelectrics. These materials are gaining attention for their multiaxial ferroelectricity, above-room-temperature Curie temperatures, and low-temperature synthesis. 
Here, we report a screening study of the Cambridge Structural Database (CSD) resulting in 55 new candidate plastic and plastic ionic ferroelectric molecular crystals, along with 16 previously reported ferroelectrics. With over $1.2~$million entries in the CSD, the screening procedure involved many steps, including considerations of molecular geometry and size, space group, and hydrogen bonding pattern.
The spontaneous polarization and electronic band gaps were predicted using density functional theory. 21 of the candidate ferroelectrics have a polarization greater than $10~\mathrm{\mu C/cm^2}$,  
out of which nine are reported at room temperature.  
\end{abstract}
 
%\keywords{Suggested keywords}%Use showkeys class option if keyword
                              %display desired
\maketitle

\section{\label{sec:level1} Introduction}

Plastic crystals are molecular crystals characterized by the existence of an
orientationally disordered mesophase.
In the mesophase, the material becomes ductile, i.e., “plastic”, due to both reduced intermolecular bonding and increased symmetry yielding facile slip planes.\cite{mondal_metal-like_2020} As a result, plastic crystals can be molded and fused, in stark contrast to most molecular crystals and inorganic ceramics, which tend to be non-flexible and/or brittle.\cite{saha_molecules_2018} 
These materials can be synthesized with low-cost and low-energy methods including co-precipitation, slow evaporation, spin coating, and 3D printing. \cite{harada_directionally_2016,deng_novel_2020,lan_cation_2021,owczarek_flexible_2016, walker_electric_2020,hu_releasing_2022}
Plastic crystals can be bonded solely through both van der Waals or hydrogen bonding and such compounds are referred to as the plastic {\it molecular} crystals, but they can also have an additional ionic component consisting of charged molecular species, which are referred to as 
plastic {\it ionic} crystals. 
Typically, the molecular species of plastic crystals have a quasi-spherical or "globular" shape, which reduces rotational barriers. \cite{timmermans_plastic_1961,ishida_structural_1989,pringle_recent_2013,wei_rational_2020,zhang_toward_2019,harada_directionally_2016,mondal_metal-like_2020} 
In addition to the plastic mesophase, the molecular rotations can imbue 
the plastic crystals with functional properties such as ferroelectricity.

Ferroelectric plastic crystals can exhibit a rich phase diagram,\cite{olejniczak_new_2013, olejniczak_pressuretemperature_2018}
with multiple competing crystalline phases and the existence of plastic ferroelectric mesophases,
as seen in quinuclidinium perrhenate, having partial orientational disorder. \cite{harada_directionally_2016}
They can also exhibit multiaxial polarization with as many as 24 equivalent axes, \cite{you_quinuclidinium_2017, xie_soft_2020, wang_room_2019, das_harnessing_2020, yang_directional_2019,harada_plasticferroelectric_2021,ai_six-fold_2020, tang_multiaxial_2017,zhang_piezoelectric_2020,zafar_raman_2019,tang_multiaxial_2018,zhang_toward_2019} 
and Curie temperatures up to 466 K.\cite{ai_highest-tc_2020, tang_organic_2020}
For applications such as FeRAM and piezoelectric sensing, a high Curie temperature is essential.  
Multiaxial polarization allows the spontaneous polarization to be aligned in a desired direction in polycrystalline systems, and can allow for multi-bit storage in single crystals. \cite{baudry_ferroelectric_2017} 
A key advantage of ferroelectric plastic crystals
is the fact that they can exhibit low coercive fields, e.g., for \ce{1-azabicyclo[2.2.1]heptanium perrhenate} and quinuclidinium perrhenate values in the $2 - 5~$kV/cm range have been reported. \cite{harada_plasticferroelectric_2019,harada_directionally_2016} Such values are comparable to that of \ce{BaTiO3}, a well-known inorganic ferroelectric.\cite{cohen_origin_1992}  
However, their spontaneous polarization tends to be on the lower end, with most reported values falling below $10~\mathrm{\mu C/cm^2}$. \cite{ harada_2018,chen_confinement-driven_2020,ai_highest-tc_2020, tang_2016, szafranski_ferroelectric_2002,budzianowski_anomalous_2008,harada_directionally_2016,you_quinuclidinium_2017,li_anomalously_2016,li_organic_2019}

In 2020, Horiuchi et al. cataloged approximately 80 reported small-molecule ferroelectric crystals. \cite{horiuchi_hydrogen-bonded_2020} This number is in stark contrast to the collection of approximately 1.2 million organic structures in the Cambridge Structural Database (CSD), \cite{cr_cambridge_2016} 
which potentially harbors many undiscovered ferroelectric plastic crystals.
Screening this database, we recently reported 6 new  organic proton-transfer ferroelectric candidates.\cite{seyedraoufi2023database}
In this paper, we detail our screening of ferroelectric candidates likely to exhibit plastic properties.
For all the identified materials, we used density functional theory (DFT) computations for geometry optimization of the crystal structure and to predict the spontaneous polarization and electronic band gaps. 
The new compounds identified in this manner are not only of interest in themselves,
but can serve as template structures for further crystal engineering, i.e., by substituting molecular species or halides to tune functional properties.

\begin{figure*}[t]
    \centering
    \includegraphics[scale=0.43]{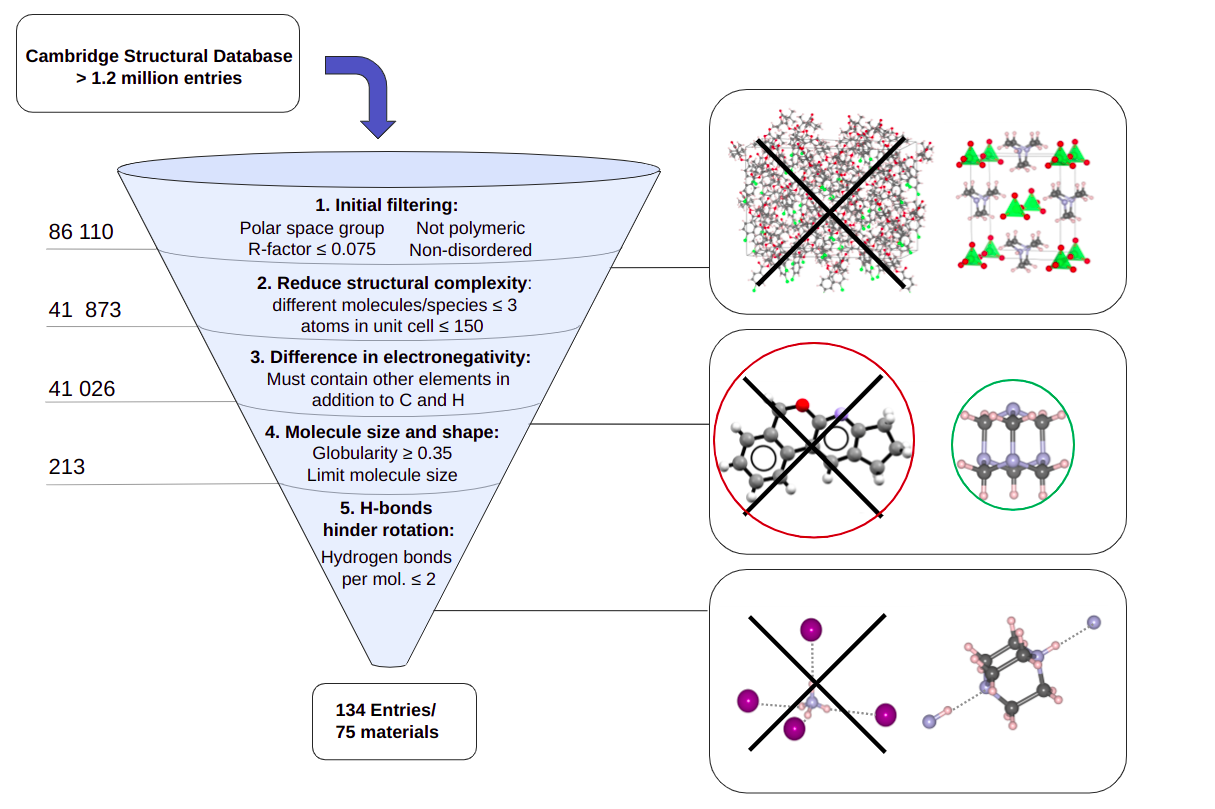}
    \caption{Overview of the screening procedure to identify ferroelectric plastic crystals in the CSD. The numbers to the left indicate the number of structures remaining after each filtering step.}
    \label{screening}
\end{figure*}

\section{Methods}
\subsection{Mining the CSD for Ferroelectric Plastic Crystals}
To identify candidate ferroelectric plastic crystals in the CSD, we 
reduced the number of structures in five filtering steps based on the properties of the molecular crystal structure and the constituents molecules,
 as outlined in Fig.~\ref{screening}.
For this procedure, the CSD Python API\cite{cr_cambridge_2016} and the \textsc{Molcrys}\cite{gitlab_2023} package developed by us, based on the Atomic Simulation Environment\cite{hjort_larsen_2017} and the \textsc{Networkx}
\cite{networkx} package were used.

\textbf{Step 1} excluded non-polar, polymeric, or disordered structures,
as well as less accurate structures, i.e., with an R-factor higher than 0.075.

\textbf{Step 2} excluded structures with unit cells containing more than $150$ atoms, removing many complex and large structures. Thus, we avoided time-consuming DFT computations.    

\textbf{Step 3} excluded all materials containing solely C and H. 
This choice was made as
polar covalent bonds or charge transfer between species is needed for high polarization, which requires electronegativity differences.  

\textbf{Step 4} excluded all structures with bulky and/or elongated molecules, as steric hindrance would typically be too large for molecular rotations in a solid phase. Only molecules with 10 or fewer non-hydrogen atoms were retained. 
Molecular graph theory was used to remove "chainy" molecules, such as
all aliphatic chains longer than four carbon atoms. This
is further detailed in Appendix \ref{sec:appendix}. 

Further, we removed structures that lacked at least one molecule with a globular or semi-globular geometry. 
As a globularity measure, we used the ratio between the volume of the convex hull and the volume of the smallest bounding sphere of the molecule, excluding hydrogen atoms.
With this measure, the \ce{C60} fullerene has a globularity of 0.87, while acetic acid has a globularity of 0.10. 
The smallest allowed globularity was set to 0.35, based on a review of known molecular ferroelectrics. 
As an example, quinuclidinium perrhenate and iodate are multiaxial ferroelectric plastic ionic crystals with low coercive fields, of respectively $340$ and $255~$kV/cm, where the quinuclidinium molecule has a
globularity of 0.52. 
In comparison, the non-plastic ferroelectric \ce{[Cu-(Hdabco)(H2O)Cl3]}\cite{zhang_ferroelectricity_2012} has a globularity of 0.24.

\textbf{Step 5} excluded structures where the globular molecules have more than two hydrogen bonds to avoid 3D hydrogen-bonded networks which would hinder molecular rotations. 

After all filters were applied, the pool was reduced to $75$ structures. 
For each, the spontaneous polarization and electronic band gap were computed using DFT.

While we identified a wide range of candidate materials,
the screening criteria have caused some candidates to be omitted.
The cap of the number of atoms in the unit cell for instance excluded the ferroelectric metal-free plastic perovskite
\ce{[NH3-dabco]NH_{4}I3}, as it has $198$ atoms in its unit cell. 
The molecular size limit furthermore
excluded some known plastic crystals, such as adamantane derivates, the \ce{C60} fullerene, \cite{andre_molecular_1992,jenkins_raman_1980,szewczyk_influence_2015}and plastic colloidal crystals.\cite{das_harnessing_2020} 
Nonetheless, our target was not to identify all plastic ferroelectrics, but rather identify several of technological interest. 
Notably, ferroelectric plastic crystals of small molecules would typically have a larger density of dipoles, both originating from individual molecules and inter-molecular charge transfer.  
Ferroelectric plastic crystals also tend to have ferroelectric phases of high symmetry, resulting in relatively small unit cells. 

Finally, this screening procedure does not evaluate if the spontaneous polarization is switchable, as is required for ferroelectrics. The presence of small, globular molecules facilitates a rotational switching mechanism. However, in some of the compounds, the switching path is not clear-cut, and more involved computations or experimental studies have to be performed for a full assessment.

\subsection{Density Functional Theory Calculations}
\label{sec:DFT}
The DFT computations were carried out using the \textsc{VASP} software package \cite{kresse_ab_1994, kresse_ab_1993, kresse_efficiency_1996, kresse_efficient_1996} with the projector augmented plane wave method (PAW) pseudopotentials. \cite{blochl_projector_1994, kresse_ultrasoft_1999} The plane wave cut-off was set to 530 eV for all computations. A $\Gamma$-centered Monkhorst-Pack \textbf{k}-point grid with a spacing of $1/15~\mathrm{\AA^{-1}}$ was used to sample the Brillouin zone. All structures were relaxed until forces fell below $0.01~\mathrm{eV/ \AA}$. The spontaneous polarization was computed using the Berry phase method. \cite{resta_macroscopic_1994, baroni_ab_1986} Both relaxation and polarization calculations were performed using the vdW-DF-cx functional, \cite{berland_exchange_2014}
which we found to provide accurate lattice constants in our benchmarking study of exchange-correlation functionals for ferroelectric plastic crystals.
\cite{sodahl_piezoelectric_2023}

\section{\label{sec:results}  Results and Discussion}

\begin{figure}[t]
    \centering
    \includegraphics[scale=0.43]{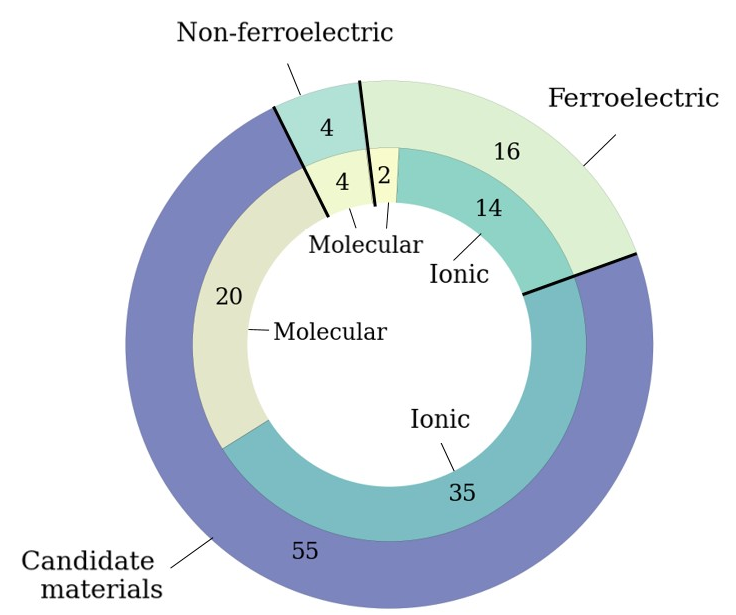}
    \caption{Overview of the candidate ferroelectrics, reported ferroelectrics, and reported non-ferroelectric materials identified in the screening. The outer circle indicates the number of structures in each of these groups, and the inner circle indicates the number of molecular and ionic molecular crystals.}
    \label{fig:mol_plastic}
\end{figure}

\begin{table*}[t]
\caption{\label{known_results} \raggedright Overview of earlier reported plastic ferroelectrics found in the CSD screening, including Curie and melting temperatures, $\mathrm{T_c}$ and $\mathrm{T_m}$  [K], coercive field $\mathrm{E_c}$ [kV/cm], experimental and computed spontaneous polarization, $\mathrm{P_{exp.}}$  and  $\mathrm{P_{calc.}}$[$\mathrm{\mu C/cm^2}$], computed electronic bandgap $\mathrm{E_g}$ [eV], crystallographic space group, alignment of dipoles, and chemical composition. Melting points are retrieved from the CSD unless a reference is listed. 
The alignment of dipoles is listed as the angle between molecular dipoles and the polarization axis, as detailed in section \ref{sec:align}.
For refcode families with more than one hit, the lowest temperature structure is listed.}
\begin{ruledtabular}
\small\begin{tabular*}{\linewidth}{l@{\extracolsep{\fill}}*{10}{c}}
CSD refcode & $\mathrm{T_c}$& $\mathrm{T_{melt.}}$ &  $\mathrm{E_c}$ & $\mathrm{P_{exp.}}$ & $\mathrm{P_{calc.}}$ & $\mathrm{E_g}$ & Spg. & Alignment &  Chem. comp.\\
\hline
\midrule
\multicolumn{10}{c}{Trimethyl-X-Y} \\
\midrule
DIRKEU01\cite{harada_2018}& 295\cite{harada_2018} & -- & 67\cite{harada_2018}& 2.0\cite{harada_2018} & 5.5 & 0.3 &Pma2 & - & \ce{(CH3)4N+}, \ce{FeCl4-}\\
\midrule
\multicolumn{10}{c}{Cyclic organic molecules} \\
\midrule
RUJBAC \cite{chen_confinement-driven_2020}& 454 \cite{chen_confinement-driven_2020}& --& -- & 1.1 \cite{chen_confinement-driven_2020}& 1.7 & 1.9 & $\mathrm{P2_1}$ & $66^{\circ}, 114^{\circ}; 36^{\circ}$ & \ce{2C6H12F2N+}, \ce{PbI4^2-} \\
BUJQIJ \cite{ai_highest-tc_2020} & 470
\cite{ai_highest-tc_2020} & -- & 10.8 \cite{ai_highest-tc_2020} & 0.48 \cite{ai_highest-tc_2020} & 2.7 & 4.6 & $\mathrm{P2_1}$ & $76^{\circ};-$ & \ce{C6H13FN+}, \ce{I-}  \\
BUJQOP \cite{ai_highest-tc_2020} & 470 \cite{ai_highest-tc_2020} & --& 9.4 \cite{ai_highest-tc_2020} & 0.40 \cite{ai_highest-tc_2020} & 2.8 & 4.6 & $\mathrm{P2_1}$ & $76^{\circ};$-& \ce{C6H13FN+}, \ce{I-}  \\
\midrule
\multicolumn{10}{c}{Dabco-based} \\
\midrule
SIWKEP \cite{szafranski_ferroelectric_2002}& 374 \cite{szafranski_ferroelectric_2002}& -- & 30 \cite{szafranski_ferroelectric_2002}& 9.0 \cite{szafranski_ferroelectric_2002}& 8.0 \cite{sodahl_piezoelectric_2023}& 3.6 & Cm & - & \ce{C6H13N2+}, \ce{ReO4-}  \\
TEDAPC28 \cite{tang_2016} & 378 \cite{olejniczak_pressuretemperature_2018}& --& -- & 4\cite{tang_2016} & 5.9 & 5.2 &$\mathrm{Pm2_1n}$ & - &\ce{C6H13N2+}, \ce{ClO4-}  \\
WOLYUR08 \cite{budzianowski_anomalous_2008}& 374 \cite{budzianowski_anomalous_2008}& -- & -- & 5\cite{budzianowski_anomalous_2008} & 6.2& 5.3 & $\mathrm{Pm2_1n}$ & - &\ce{C6H13N2+}, \ce{BF4-}  \\
BILNES \cite{ye_metal-free_2018} & 390 \cite{ye_metal-free_2018} & -- & -- & --& 15.9 & 4.5 &R3 & $52^{\circ};-$ & \ce{C7H16N2^2+}, \ce{NH4+}, \ce{3Br-} \\
BILNOC \cite{ye_metal-free_2018} & 446 \cite{ye_metal-free_2018} & -- & 6-12\cite{ye_metal-free_2018} & 22\cite{ye_metal-free_2018} & 21.9 & 4.0 &R3 & $52^{\circ};-$ & \ce{C7H16N2^2+}, \ce{NH4+}, \ce{3I-}  \\
\midrule
\multicolumn{10}{c}{Quinuclidinium-based} \\
\midrule
LOLHIG02 \cite{tang_organic_2020} & 466 \cite{tang_organic_2020} & --& -- & 11.4 \cite{tang_organic_2020} & 12.7 \cite{sodahl_piezoelectric_2023} & 4.4 & Pn & $3^{\circ};-$ & \ce{C7H13FN+}, \ce{ReO4-}  \\
OROWAV \cite{harada_directionally_2016} & 367 \cite{harada_directionally_2016} & -- & 340 \cite{harada_directionally_2016}& 5.2 \cite{harada_directionally_2016}& 7.3 \cite{sodahl_piezoelectric_2023}& 4.5 &$\mathrm{Pmn2_1}$ & $5^{\circ};-$ & \ce{C7H14N+}, \ce{ReO4-}  \\
YASKIP\cite{you_quinuclidinium_2017} & 322\cite{you_quinuclidinium_2017} & -- & 255\cite{you_quinuclidinium_2017} & 6.7\cite{you_quinuclidinium_2017} & 6.5 \cite{sodahl_piezoelectric_2023}& 2.9 &$\mathrm{Pmn2_1}$ & $6^{\circ};-$ & \ce{C7H14N+},\ce{IO4-}  \\
SIYWUT\cite{Siczek_2008, li_anomalously_2016} & -- & -- & 1000\cite{li_anomalously_2016} & 1.7\cite{li_anomalously_2016} & 5.6 \cite{sodahl_piezoelectric_2023}& 5.3 &$\mathrm{P4_1}$ &$113^{\circ};-$& \ce{C7H14NO+}, \ce{Cl-} \\
ABIQOU\cite{li_anomalously_2016} & -- & -- &-- & -- & 5.2 \cite{sodahl_piezoelectric_2023} & 5.0 &$\mathrm{P4_1}$ & $113^{\circ};-$& \ce{C7H14NO+}, \ce{Br-}  \\
MIHTEE \cite{li_organic_2019}& 400\cite{li_organic_2019} & 492 & -- & 6.96\cite{li_organic_2019} & 10.2 & 4.6 &$\mathrm{P6_1}$ & $85^{\circ} $&\ce{(R)-C7H13NO}  \\
QIVQIY \cite{li_organic_2019}& 400\cite{li_organic_2019} & -- & -- & 6.72\cite{li_organic_2019} & 10.2 & 4.5 &$\mathrm{P6_5}$ & $85^{\circ} $& \ce{(S)-C7H13NO}  \\
\midrule
\multicolumn{10}{c}{Reported non-ferroelectrics } \\
\midrule
BOXCUO \cite{shi_novel_2014}&-- &--&-- & -- & -- & 0 &$\mathrm{P6_3mc}$ & - &\ce{C4H12P+}, \ce{FeCl4-}  \\
QIMXER\cite{Szafrański_2013}&--&--&-- &  -- & 9.1& 4.7 &  $\mathrm{Cmc2_1}$ & - &\ce{C6H13N2+}, \ce{Cl-}  \\
BOBVIY12\cite{Olejniczak_2010}&--&--&-- &  -- & 6.7 & 3.8 & $\mathrm{Pmc2_1}$ &- &\ce{C6H13N2+}, \ce{I-}  \\
BOCKEK06\cite{szafranski_2008}&--&--&--& -- & 12.2 & 4.8 &$\mathrm{P1}$ & - & \ce{C6H13N2+}, \ce{HF2-}  \\
\end{tabular*}
\end{ruledtabular}
\end{table*}

\begin{table*}
\caption{\label{candidate_results1} Overview of candidate ferroelectric plastic crystals discovered in the screening of the CSD, including melting temperatures $\mathrm{T_{melt.}}$ [K], computed spontaneous polarization $\mathrm{P_{calc.}}$ [$\mathrm{\mu C/cm^2}$], computed electronic bandgap $\mathrm{E_g}$ [eV], crystallographic space group, alignment of dipoles and chemical composition. Melting points are retrieved from the CSD unless a reference is listed.   $\mathrm{T_{struct.}}$ is the temperature the structure is measured as, (RT) indicates that the structure is also reported at room temperature.
The alignment of dipoles is listed as the angle between molecular dipoles and the polarization axis, as detailed in section \ref{sec:align}. For refcode families with more than one hit, the lowest temperature structure is listed.}
\begin{ruledtabular}
\begin{tabular}{*9c}
&CSD refcode & $\mathrm{T_{struct.}}$ & $\mathrm{T_{melt.}}$ & $\mathrm{P_{calc.}}$ & $\mathrm{E_g}$ & Spg. & Alignment & Chem. comp.\\
\hline
\midrule
\multicolumn{9}{c}{Trimethyl-X-Y} \\
\midrule
\midrule
{\multirow{7}{*}{\rotatebox[origin=c]{90}{M(olecular)}}}
&ZZZVPQ01 & RT &  -- & 3.0 & 5.8 &$\mathrm{R3mr}$ & $0^{\circ}$ & \ce{(CH3)3SO3N}  \\
&ZZZVPE02& 150 (RT) & -- &  5.4 & 6.4 &R3m & $0^{\circ}$ & \ce{(CH3)3BH3N}  \\
&LINZOX & RT & -- & 6.4& 4.4 & $\mathrm{Ama2}$ & $63^{\circ} $ & \ce{(CH3)3NH3Al}  \\
&TMAMBF11 & 100 (RT) & 367\cite{mondal_metal-like_2020}  & 20.3 & 7.4 &R3m & $0^{\circ}$& \ce{(CH3)3BF3N}  \\
&CAVJOZ & 193& --  &0.04 & 2.8 &$\mathrm{P6_2mc}$ & - &\ce{(CH3)3Cl2Nb}  \\
&TBUHLB05 & 180 &  254  & 7.3 & 4.9 & $\mathrm{Pmn2_1}$ & $19^{\circ}, 32^{\circ}, 33^{\circ} $ & \ce{(CH3)3BrC}  \\
&ETIPAY & 160&  183 &  5.5 & 5.3 &$\mathrm{Pna2_1}$ &  $20^{\circ}$ & \ce{(CH3)3CH2ClSi} \\
\midrule
{\multirow{15}{*}{\rotatebox[origin=c]{90}{I(onic)}}}
&WAGGAM & RT &  -- & 11.0 & 2.7 &$\mathrm{P2_1}$ &$6^{\circ}, 83^{\circ};$- & \ce{2(CH3)3OS+}, \ce{Cr2O7^2-}  \\
&GEPZIK & 123&  185 & 3.9 &7.6 &$\mathrm{P6_3}$ & - & \ce{(CH3)3HN+}, \ce{F-}, \ce{6HF}  \\
&ZISCUC & 300 & -- &  14.4 & 0.2 &$\mathrm{Cm}$ & $70^{\circ};$- & \ce{(CH3)3CH2ClN+}, \ce{FeCl4-} \\
&YODGON & RT &-- & 3.1 & 5.0 & $\mathrm{Pmn2_1}$ & - & \ce{(CH3)4N+}, \ce{OCN-}  \\
&XAKBUG & 223 &  -- & 15.3& 2.6 &Abm2 & $103^{\circ};$- & \ce{(CH3)4N+}, \ce{OsFO4-}  \\
&VUGNUG & RT &  -- & 22.0 & 5.3 &$\mathrm{Pmn2_1}$ & - & \ce{(CH3)4N+}, \ce{N3-}  \\
&MIWBEC & 293 &  -- & 13.4 & 0.2 &$\mathrm{Pca2_1}$ & $53^{\circ};$- &\ce{(CH3)4N+}, \ce{FeCl3NO-}  \\
&ORUKUK & 100 & -- & 12.9 & 3.0 &$\mathrm{Pna2_1}$ & - & \ce{(CH3)4N+}, \ce{Cl3F4-} \\
&ZOYGUP & RT &  484  & 3.1 & 4.1 &$\mathrm{P2_1}$ &$68^{\circ};-$&\ce{(CH3)4N+}, \ce{C5H7O4-}  \\
&SEYLAJ & 123 & 317  & 12.1 & 5.2 &$\mathrm{P3_1}$ & - & \ce{(CH3)4N+}, \ce{OH-}, \ce{4H2O}  \\
&PEVXOE & 100&   -- & -- & 0 &$\mathrm{Cmc2_1}$ & -  &\ce{(CH3)4P+}, \ce{O2-}, \ce{2NH3}  \\
&PEVXUK & 100 & -- & -- & 0 &$\mathrm{Cmc2_1}$ & - & \ce{(CH3)4As+}, \ce{O2-}, \ce{2NH3}  \\
&XENFAZ & 295&  493  & 8.4 & 1.7 &$\mathrm{P2_1}$& $17^{\circ}, 161^{\circ};$-&\ce{2(CH2O)3NH3C+}, \ce{HgI4^2-}  \\
\midrule
\midrule
\multicolumn{9}{c}{Dabco-based} \\
\midrule
\midrule
{\multirow{2}{*}{\rotatebox[origin=c]{90}{M}}}
&LOLWEO & RT & 429 & 7.6 &3.4 &R3m &  $18^{\circ},161^{\circ};$- & \ce{C6H12N2}, \ce{2CH4N2S}  \\
&HUSRES & 150& -- & 8.2 & 4.8 &$\mathrm{Cc}$ & -$;14^{\circ}$ &\ce{C6H12N2}, \ce{C2H5O5P}  \\
\midrule
{\multirow{4}{*}{\rotatebox[origin=c]{90}{I}}}
&USAFIG & 295&  -- & 0.4 & 3.4&$\mathrm{Pna2_1}$ & $95^{\circ} ;$-& \ce{C6H13N2O2+}, \ce{NO3-}  \\
&VAGVAA01 & 150&  -- & 12.8& 5.1 &$\mathrm{Pna2_1}$ & - & \ce{C6H14N2^2+}, \ce{2Cl-}  \\
&GASBIO & 123 & -- & 11.2 & 4.2 &$\mathrm{Pca2_1}$ & - & \ce{C6H14N2^2+}, \ce{2I-}, \ce{H2O} \\
&NAKNOF03 & 150& -- & 18.4 & 6.8 &P1 &-&\ce{C6H14N2^2+}, \ce{2BF4-}, \ce{H2O}  \\
\midrule
\midrule
\multicolumn{9}{c}{Hexamine-based} \\
\midrule
\midrule
{\multirow{1}{*}{\rotatebox[origin=c]{90}{M}}}
&INEYUY/TAZPAD & 100 (RT) &  -- & 0.9 & 3.8 &R3m & $0^{\circ}$& \ce{C6H12N3P}  \\
\midrule
{\multirow{3}{*}{\rotatebox[origin=c]{90}{I}}}
&HMTAAB & RT & -- & 11.6 & 4.8 &$\mathrm{P6_3mc}$ & - & \ce{C6H12N4}, \ce{NH4+}, \ce{BF4-} \\
&BOHNUH01 & 295&   -- & 10.6 & 4.9 &R3m & $83^{\circ};$-& \ce{C6H13N4+}, \ce{Br-}  \\
&TOZTAF & 296 & -- & 17.3 & 4.3 &$\mathrm{Cc}$ & $138^{\circ};89^{\circ}$&\ce{C6H13N4+}, \ce{C4H5O5-}  \\
\midrule
\midrule
 \multicolumn{9}{c}{Boron clusters} \\
\midrule
\midrule
{\multirow{2}{*}{\rotatebox[origin=c]{90}{M}}}
&SASSOU & RT & 378 & 8.7 & 3.5 &Cc& $80^{\circ}$ &\ce{CH10B6S2} \\
&OTOLAM & 150 &  -- & 11.5 & 5.5 &Pnn2 & $0^{\circ}$ & \ce{C2H14B8}  \\
\midrule
{\multirow{2}{*}{\rotatebox[origin=c]{90}{I}}}
&UTUZAM & 90 & -- & 3.2& 4.8 &$\mathrm{P2_1}$ & $56^{\circ};$- & \ce{CH14B9-}, \ce{C3H10N+}\\
&LUWHOD & 340 & -- & 7.8 & 4.4 &$\mathrm{P4_2}$& - &\ce{B10H10^2-}, \ce{2NH4+}, \ce{NH3} \\
\midrule
\midrule
\multicolumn{9}{c}{Cyclic organic molecules} \\
\midrule
\midrule
{\multirow{6}{*}{\rotatebox[origin=c]{90}{I}}}
&WAQBOH & 295 & -- & 8.1& 5.7 &Pn & - & \ce{C4H11N2+}, \ce{BF4-}  \\
&CUWZOM & 100&  -- & 0.4& 5.5 &$\mathrm{P2_1}$&$52^{\circ};$- & \ce{C5H11FN+}, \ce{Cl-}  \\
&AMINIT & 293 & -- &6.8 & 4.5& $\mathrm{P2_1}$ & $126^{\circ}; 88^{\circ} $ & \ce{C5H12N+}, \ce{H2AsO4-} \\
&EVULAI & 100 & -- & 13.2 & 4.7 &$\mathrm{P2_1}$ & $39^{\circ}$& \ce{C6H15N2S+}, \ce{Cl-} \\
&FENYEC & RT & 407 & 4.7 & 3.5 &$\mathrm{Cc}$ & $55^{\circ}; 85^{\circ}$ &\ce{C7H5O3-}, \ce{C4H10N+}  \\
&HAJXEW & 123 & -- & 16.5 & 5.6 &Pn & $16^{\circ};$-&\ce{C6H18N3^3+}, \ce{ClO4-}, \ce{2Cl-}\\
&OBEXEY &  200 & 416 & 7.6 & 4.8 &$\mathrm{Cmc2_1}$ & - & \ce{C9H20N+}, \ce{N3-}  \\
\end{tabular}
\end{ruledtabular}
\end{table*}

\begin{table*}
\ContinuedFloat
\caption{\label{candidate_results2} Continuation of Table~\ref{candidate_results1}}
\begin{ruledtabular}
\begin{tabular}{*9c}
&CSD refcode& $\mathrm{T_{struct.}}$ & $\mathrm{T_{melt.}}$ & $\mathrm{P_{calc.}}$ & $\mathrm{E_g}$ & Spg. & Alignment &  Chem. comp.\\
\midrule
\midrule
\multicolumn{9}{c}{Cage-like organic molecules} \\
\midrule
\midrule
{\multirow{8}{*}{\rotatebox[origin=c]{90}{M}}}
&EQAXOL & 140&  -- & 9.1 &5.4 & $\mathrm{Pna2_1}$ & $48^{\circ}$& \ce{C5H9O3P} \\
&BAPFAB & 150 & -- & 4.1 & 5.3 &$\mathrm{P2_1}$ & $20^{\circ}, 56^{\circ}$ & \ce{C6H7FO3} \\
&BAPFOP & 150& -- & 0.7 & 5.4 &$\mathrm{P2_1}$ & $20^{\circ}$ & \ce{C6H7FO3}\\
&NOCPIE01 & 100 (RT) &  -- & 14.9 & 4.5 &$\mathrm{P2_1}$ & $51^{\circ},63^{\circ} $ &\ce{C7H10O3}  \\
&MIRHUQ &180 &388 (sublim.) & 11.3& 4.6 &$\mathrm{P2_1}$ &$29^{\circ}, 31^{\circ}$  & \ce{C8H15PS} \\
&XIBVIN & 180&  443 & 1.3 & 4.2 &$\mathrm{P2_1}$ & $77^{\circ}$&\ce{C8H15PS} \\
&BESWON & 130 &-- & 4.5 & 5.7 &$\mathrm{P3_1}$ & $70^{\circ}$ & \ce{C9H16O}  \\
&FEJFAB & RT&  -- & 7.4 & 2.1&$\mathrm{Pmc2_1}$ & $34^{\circ};$- &  \ce{C6H10N2}, \ce{CBr4}\\
\midrule
{\multirow{3}{*}{\rotatebox[origin=c]{90}{I}}}
&PINRAI & 173&  -- & 5.3 & 5.8 &$\mathrm{Pna2_1}$ & $60^{\circ};$- & \ce{C6H9F3N+}, \ce{Cl-}  \\
&QAZFUV & 100&  -- & 11.5 & 5.2 &$\mathrm{Cmc2_1}$ & $54^{\circ};$-& \ce{C6H12N+}, \ce{Cl-}  \\
&JEBVOC & 293&  $>533$  &8.2 & 4.4&$\mathrm{P2_1}$ & $70^{\circ};$-& \ce{C9H14N+}, \ce{Cl-} \\
\midrule
\midrule
\multicolumn{9}{c}{Other} \\
\midrule
\midrule
{\multirow{3}{*}{\rotatebox[origin=c]{90}{I}}}
&HUPTUI & 100 &-- & 0.4 & 3.7 &$\mathrm{P2_1}$ & $86^{\circ};-$& \ce{C6H18N3S+}, \ce{TaF6-} \\
&VAJKUM & RT & 513  & 1.7 & 3.2 &$\mathrm{P2_1}$ & $ 85^{\circ}; 43^{\circ}$ &\ce{C6H18N3S+}, \ce{SF5O-} \\
&OTIJUZ & 123& 393 \cite{Bläsing_2021}  & 2.6 & 5.1 &Pc& $43^{\circ}, 120^{\circ};-$ &\ce{2C6H18OPSi+}, \ce{S2O7^2-}\\
\end{tabular}
\end{ruledtabular}
\end{table*}

Out of the 75 materials identified by screening the CSD, 16 are earlier reported to have ferroelectric properties. \cite{harada_2018,chen_confinement-driven_2020,ai_highest-tc_2020,szafranski_ferroelectric_2002,tang_2016,olejniczak_pressuretemperature_2018,budzianowski_anomalous_2008,ye_metal-free_2018,tang_organic_2020,harada_directionally_2016,you_quinuclidinium_2017, Siczek_2008, li_anomalously_2016,li_organic_2019} Four have also been studied for their %in relation to their ferroelectric, 
piezoelectric and/or dielectric properties, but were not reported to be ferroelectric (CSD refcodes: BOXCUO, QIMXER, BOBVIY12, and BOCKEK06).\cite{shi_novel_2014, Szafrański_2013,Budzianowski_2006, Olejniczak_2010,szafranski_2008} Table~\ref{known_results} lists the available experimental results, as well as the computed spontaneous polarization and electronic band gaps for the earlier reported ferroelectrics and non-ferroelectrics, while Table~\ref{candidate_results2} lists the computed values for the candidate materials.
For three of them, the DFT computations did not yield a band gap. This was confirmed using the hybrid functional 
of Heyd, Scuseria, and Ernzerhof, HSE06,\cite{krukau_influence_2006},
which does not underestimate band gaps as the 
vdW-DF-cx functional does due to the lack of non-local exchange. 

Fig.~\ref{fig:mol_plastic} shows an overview of the molecular and ionic molecular crystals among the earlier reported and candidate ferroelectrics. Out of the 16 known ferroelectrics, 14 are ionic molecular crystals. For the candidate materials, 20 are molecular, and 35 are ionic molecular crystals. 

\begin{figure*}[t]
    \centering
    \includegraphics[scale=0.72]{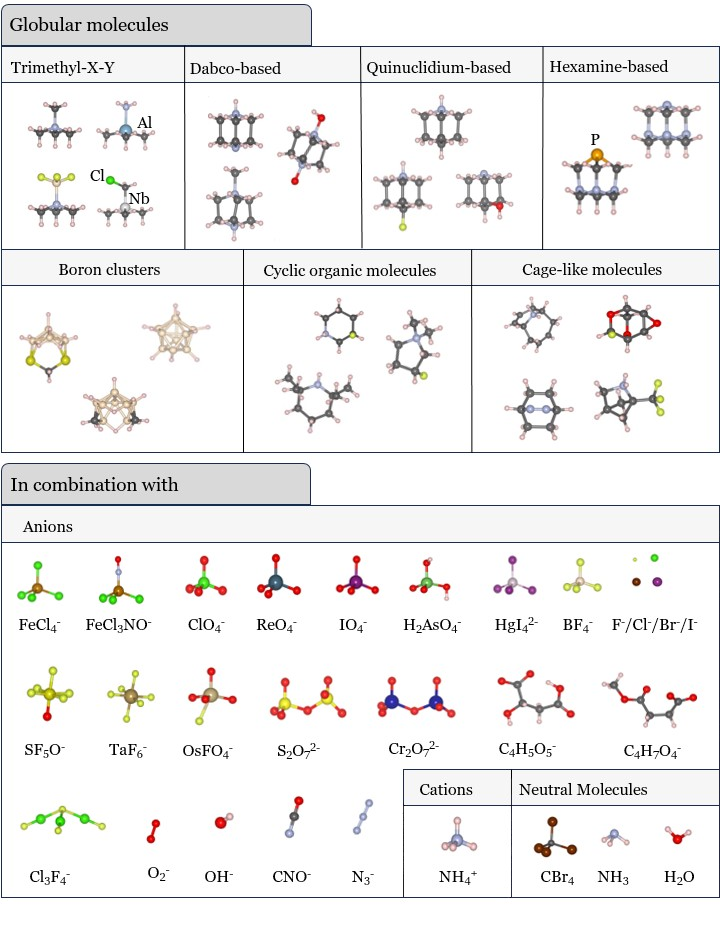}
    \caption{Examples of globular molecules and the anions, cations, and neutral molecules that are combined in the identified structures. In the top panel, carbon atoms are shown as gray, nitrogen as blue, oxygen as red, fluorine as yellow, boron in beige, and hydrogens as white. Other elements are labeled.}
    \label{fig:mols}
\end{figure*}

A selection of the chemical species found in the identified materials, including the globular molecules and the neutral molecules and ions combined with the globular is illustrated in Fig.~\ref{fig:mols}.
In total, the materials contain 30 different anionic molecules, 31 cations, and 25 neutral molecules.

\begin{figure*}[t]
    \centering
    \includegraphics[scale=0.25]{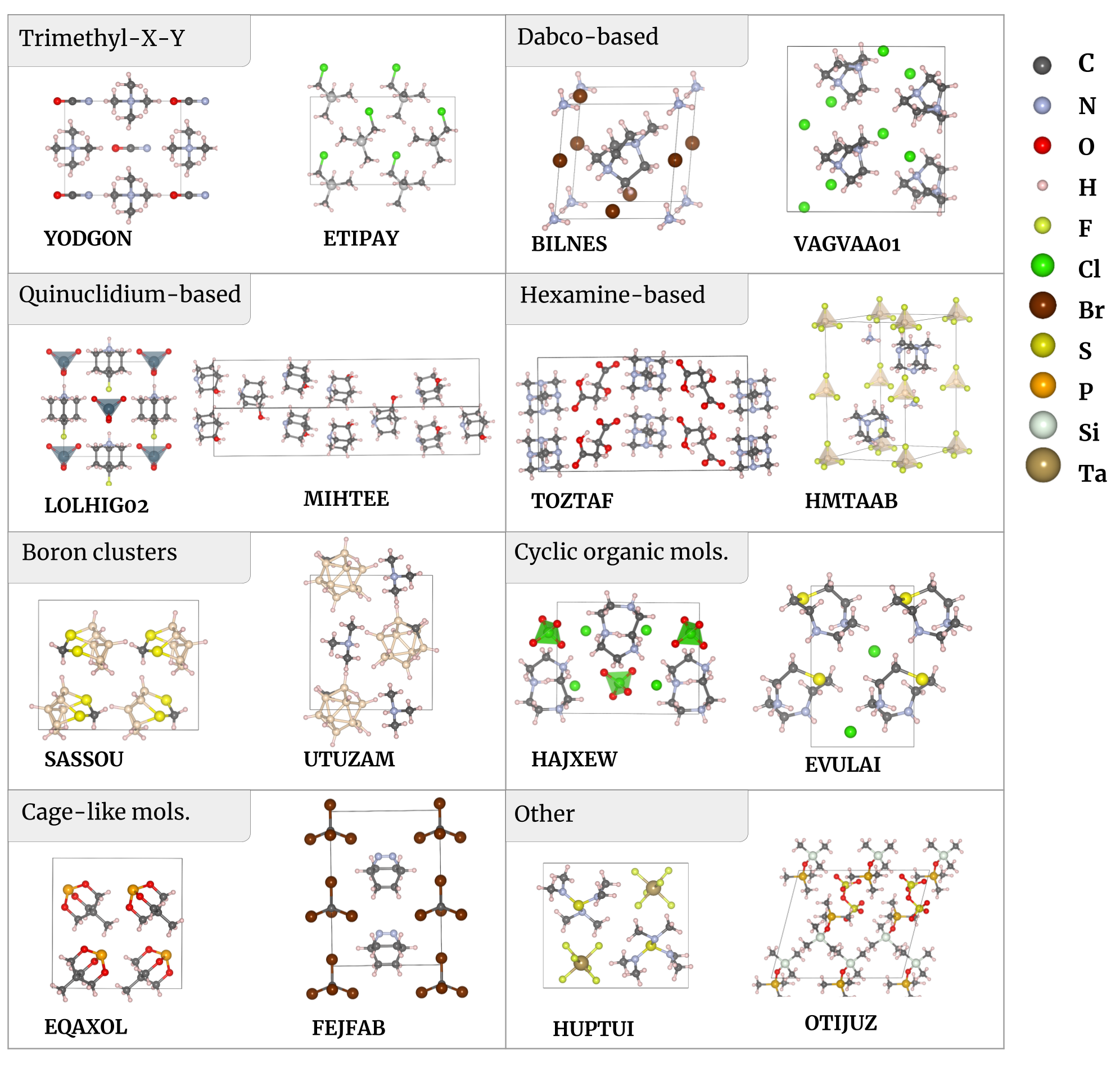}
    \caption{Examples of structures found in each of the eight groups of materials. ETIPAY, MIHTEE, SASSOU, EQAXOL, and FEJFAB are molecular crystals, the remaining are ionic.}
    \label{fig:all_struct}
\end{figure*}

Fig.~\ref{fig:all_struct} illustrates the seven groups of materials identified, based on the composition and geometry of the globular molecules. 
Three candidate materials do not fit into any of these categories, they are listed as "Other" in Table~\ref{candidate_results2}. 

\begin{figure}
    \centering
    \includegraphics[scale=0.51]{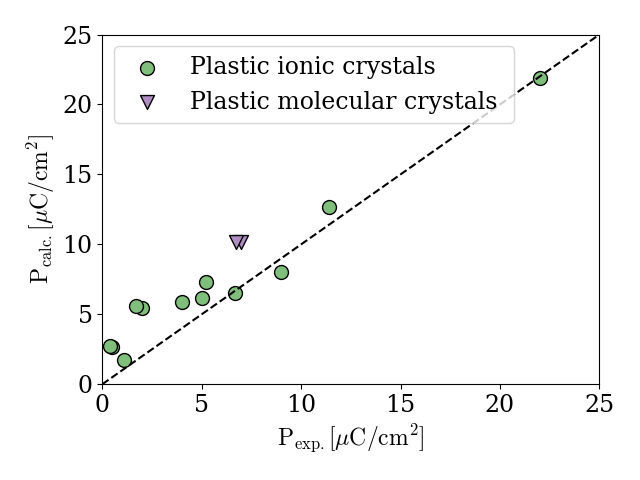}
    \caption{\raggedright The computed spontaneous polarization compared with the experimentally measured values. 
    Spontaneous polarization is reported for 14 out of the 16 known ferroelectrics identified in the screening.
    }
    \label{fig:P_compare}
\end{figure}

Fig.~\ref{fig:P_compare} shows that the 
experimentally measured spontaneous polarizations\cite{harada_2018, chen_confinement-driven_2020,ai_highest-tc_2020,szafranski_ferroelectric_2002,tang_2016,budzianowski_anomalous_2008,ye_metal-free_2018,tang_organic_2020,harada_directionally_2016,you_quinuclidinium_2017,li_anomalously_2016,li_organic_2019}  agree well with previously reported values for plastic molecular and plastic ionic molecular crystals,
but computed values are often slightly higher than the experimental.
Extrinsic effects such as defects, grain orientation and boundaries, and electronic leakage
can reduce the experimental spontaneous polarization.\cite{maglione_defect_2015,damjanovic_ferroelectric_1998}
Moreover, atomic vibration and molecular librations can also reduce the measured spontaneous polarization.
In assessing the properties with DFT, 
we used the lowest temperature ferroelectric phase reported in CSD,
as the dynamic molecular motion could play a significant role at elevated temperatures and in particular in the high-temperature phases.
Note that for 44 of the materials, the structure reported at the lowest temperature coincides with the room temperature phase. For four materials, a paraelectric phase is reported at room temperature, and for one material, VAGVAA01, the room temperature structure is a different potentially ferroelectric phase.
Finally, for 26 of the materials, no crystalline room temperature phase has been reported in the CSD. 

\subsection{Candidate Ferroelectric Plastic Crystals}

\subsubsection{Trimethyl-X-Y materials}

One group of plastic ferroelectric candidates is the 
trimethyl-X-Y materials, where X denotes an atom and Y 
a chemical group or atom; two structures illustrated in Fig.~\ref{fig:all_struct}.
Out of the 21 materials, only one (DIRKEU01)
has previously been reported as ferroelectric.
Of the remaining 20, seven are based on neutral molecules, while the rest are constructed from various combinations of 12 anions and seven different cations, the most common being tetramethylammonium, found in 8 out of the 21 materials. 
This group of materials shows significant promise, with eight having computed spontaneous polarization values exceeding $10~\mathrm{\mu C/cm^2}$ (Table~\ref{candidate_results1}). 
The largest value is found for VUGNUG with $22~\mathrm{\mu C/cm^2}$. 
For two of the materials, PEVXOE and PEVXUK, we find no bandgap and therefore no spontaneous polarization is listed. 

\subsubsection{Dabco-based Materials}
Materials constructed from various derivatives of \ce{1,4-diazabicyclo[2.2.2]octane} (dabco) in combinations with anions such as \ce{ClO4-} and \ce{ReO4-} 
are attractive ferroelectrics with
low coercive fields, and rapid ferroelectric switching reported with frequencies up to $263~$kHz.\cite{sun_molecular_2018,tang_ultrafast_2016, fu_high-tc_2020,szafranski_ferroelectric_2002,tang_multiaxial_2017} 
Moreover, Curie temperatures as high as $540~$K
have been reported by Li et al.\cite{li_multiaxial_2021} 

In our study, we found 11 dabco-based materials, out of which five have previously been reported as ferroelectrics.\cite{szafranski_ferroelectric_2002,tang_2016,olejniczak_pressuretemperature_2018,budzianowski_anomalous_2008,ye_metal-free_2018}
Two of these, BILNES and BILNOC, are organic metal-free perovskites, \cite{ye_metal-free_2018} see Fig.~\ref{fig:all_struct}.
Interestingly, two of the candidate materials, LOLWEO and HUSRES, are co-crystals of charge-neutral molecules.

The computed spontaneous polarization values for the known ferroelectric dabco-based materials range from $\mathrm{3.56}$ to $ {21.9~\mathrm{\mu C/cm^2}}$ for BILNOC,
compared to  $\mathrm{0.4 - 18.4~\mu C/cm^2}$ for the identified candidate ferroelectric plastic crystals. 

\subsubsection{Quinuclidinium-based Materials}
Several materials containing variations of the quinuclidinium molecule
have been studied in recent years. \cite{Siczek_2008,li_anomalously_2016, li_organic_2019,yoneya_molecular_2020,deng_novel_2020,lee_stabilization_2021, yang_directional_2019,xie_soft_2020, zhang_toward_2019}
Notably, Tang et al. reported 
a Curie temperature of $466~$K\cite{tang_organic_2020} for \ce{[F-C7H13N]ReO4}.
Another compound of interest is \ce{[C7H14N]IO4}, for which 
You et al. found 12 equivalent directions of polarization.\cite{you_quinuclidinium_2017} 
The reported coercive fields vary from $\mathrm{255}$ all the way up to $\mathrm{1000~kV/cm}$.\cite{li_anomalously_2016} 
All the seven quinuclidinium-based materials found in our study have previously been reported as ferroelectrics, \cite{tang_organic_2020,harada_directionally_2016,you_quinuclidinium_2017,harada_plasticferroelectric_2019,Siczek_2008, li_anomalously_2016,li_organic_2019} (Table \ref{known_results}). 
Five of them are plastic ionic crystals, while the last two are plastic molecular crystals and stereoisomers of the same compound. 
The computed spontaneous polarizations of the  quinuclidinium materials range 
from $5.2$ to $12.7~\mathrm{\mu C/cm^2}$.

\subsubsection{Hexamine-based Materials}
Four hexamine-based crystal structures were identified in the screening study,
none of which (to our knowledge) have been reported as ferroelectrics in the past.
Fig.~\ref{fig:mols} shows the two variations of the hexamine molecule found, the regular hexamine molecule, and one where a nitrogen atom has been substituted by phosphorus. Three structures are ionic molecular crystals, with spontaneous polarizations ranging from $10.6$ to $18.4~\mathrm{\mu C/cm^2}$ (Table~\ref{candidate_results1}). 
One of these, HMTAAB, has a 
perovskite-like structure, see Fig.~\ref{fig:all_struct}.
The non-ionic molecular crystal with a spontaneous polarization of $0.9~\mathrm{\mu C/cm^2}$ is reported with two refcodes in the CSD, TAZPAD and INEYUY. 
The large spontaneous polarization values of the ionic hexamine-based  materials should encourage further experimental characterization and optimization of this group of materials.

\subsubsection{Boron Cluster Materials}
While boron clusters have been studied for their characteristic chemistry, biological application, as well as magnetic, optic, and electronic properties, \cite{li_iron_2009, ma_pben05b5o8oh_2015,zhao_density_2021,barba-bon_boron_2022} their ferroelectric properties have not received particular attention. 
Our screening study identified four materials containing boron clusters (Table~\ref{candidate_results1}). Two, SASSOU and LUWHOD, are reported at room temperature or higher and have spontaneous polarizations around  $8~\mathrm{\mu C/cm^2}$. The highest polarization is computed for OTOLAM, at $11.5~\mathrm{\mu C/cm^2}$.
The relatively large polarization combined with room temperature stability makes the boron cluster-based materials interesting for further studies.

\subsubsection{Materials Based on Cyclic Organic Molecules}
Some of the crystal structures identified contain cyclic organic molecules, both aromatic or non-aromatic, 
and we grouped these together. 
The 10 materials identified are all ionic molecular crystals including three known ferroelectrics \cite{chen_confinement-driven_2020,ai_highest-tc_2020} (Table \ref{known_results}), two of them being stereoisomers of the same compound. The third, RUJBAC, is an organic-inorganic hybrid perovskite. \cite{chen_confinement-driven_2020} 
For all three, the spontaneous polarization fall below $3~\mathrm{\mu C/cm^2}$. 
The coercive fields are in the range $\mathrm{10~kV/cm}$,\cite{ai_highest-tc_2020}, and they all exhibit Curie temperature exceeding $450~$K.\cite{ai_highest-tc_2020,chen_confinement-driven_2020}
Among the seven candidates identified here,
all but one have spontaneous polarizations exceeding the earlier reported ferroelectrics, HAJXEW having the largest value of $16.5~\mathrm{ \mu C/cm^2}$ (Table \ref{candidate_results1}).
Three of the candidates identified, WAQBOH, AMINIT, and FENYEC have been reported as stable at room temperature, with the latter melting above $400~$K. 

\subsubsection{Materials Based on Organic Cage-like Molecules}
We found eleven crystals structures with various 
organic cage-like molecules, none of which have previously been reported as ferroelectric. 
Seven of these are molecular crystals, while four are ionic (Table~\ref{candidate_results2}).  
Even though there are no reported room temperature structures for MIRHUQ and XIBVIN, their reported sublimation and melting temperatures are high, $388$ and $443~$K, respectively. Interestingly, these two structures are polymorphs of the same compound. JEBVOC is reported to have a melting temperature larger than $533~$K.

Finally, three of the candidate ferroelectrics do not fit into any of the defined groups of materials. The computed spontaneous polarizations are below $3~\mathrm{\mu C/cm^2}$ for all of them. HUPTUI and VAJKUM are isostructural and are built up of tris(dimethylamino)sulfonium molecule and an octahedral inorganic anion. 

\subsection{Comparison of Candidate and Previously Reported Ferroelectrics}

The previously reported ferroelectrics frequently contain cage-like organic molecules, such as dabco and  quinuclidinium-derivatives, combined with halogen, \ce{FeCl4-} or \ce{XO4-} anions. The molecules in the
identified candidates are by comparison generally smaller: 25 materials consist of molecules with five or fewer carbon atoms (not counting structures of boron clusters). The smaller volume of the asymmetric unit allows for larger spontaneous polarization. For example, DIRKEU01 is a known ferroelectric with a computed polarization of $5.5~\mathrm{\mu C/cm^2}$, VUGNUG is a candidate material with a value of $22.0\mathrm{\mu C/cm^2}$. Both materials are of the trimethyl-X-Y group with 0.5 formula units per asymmetric unit, and the cell volume of DIRKEU01 is almost twice as large as VUGNUG. The smaller volume of VUGNUG results in a higher dipole density, and thus the larger spontaneous polarization.
The lower molecular weights of the smaller molecules of the candidate materials could, however, lead to reduced melting points. Overall, the candidates display a more diverse set of anions that include entities like \ce{HF2-}, \ce{CNO-}, \ce{FeCl3NO-}, and \ce{SF5O-}.

\subsection{\label{sec:align}Polarization and Alignment of Molecules}

Molecules in plastic crystals often pack in complex arrengements, and the direction of the individual molecular dipoles does not necessarily align with the direction of the spontaneous polarization.\cite{sodahl_piezoelectric_2023} 
In Table \ref{known_results} and \ref{candidate_results1}, 
the "dipolar" direction relative to the polarization direction is listed. 
Typically, the unit cell holds several equivalent molecules that align at the same angle with the polarization axis, as given by the space group symmetry.
Mirror and/or rotational symmetries cause the polarization contributions perpendicular to the polarization axis to cancel out.
For many of the molecules, the dipolar direction is evident from their symmetry,
but not all molecules have a clear direction. 
We therefore conveniently obtain the "dipole" direction from the moments of the  electronegativity relative to the center of electronegativity.

\begin{figure}[t]
    \centering
    \includegraphics[scale=0.37]{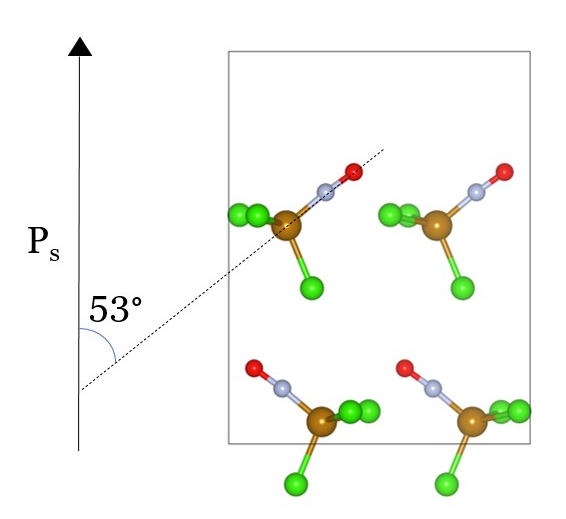}
    \caption{Illustration of the alignment of dipoles in MIWBEC (tetramethylamine molecules omitted for simplicity). The dipoles align along two directions, both $53^{\circ}$ on the polarization axis.}
    \label{fig:alignment}
\end{figure}

Fig.~\ref{fig:alignment} illustrates the alignment of the molecular dipoles relative to the polarization axis for MIWBEC. 
In the cases where the molecules are symmetrical or have a negligible dipole, no alignment is listed. In these materials, interspecies charge transfer is the dominant contribution to the spontaneous polarization. 
Examples of such systems, are six of the dabco-based materials, SIWKEP, TEDAPC28, WOLYUR08, VAGVAA01, GASBIO, and NAKNOF03.
Despite negligible molecular dipoles, their spontaneous polarizations 
are in the range $5.0-18.4~\mathrm{\mu C/cm^2}$. 
For systems where the molecular dipoles run counter to the overall polarization, an interesting prospect opens up for realignment of the dipoles by applying an electric field. This can both increase the spontaneous polarization and allow for multi-bit storage, assuming that the electric field required to realign the dipoles is smaller than the coercive field of the ferroelectric material.

\subsection{Plastic Properties of the Candidate Ferroelectrics}

While the screening study can identify candidate ferroelectric plastic crystals, truly predicting whether the materials can transition to a plastic mesophase demands more involved simulations such as molecular dynamics. This is not feasible for the large pool of candidate materials identified in this study, but out of the 16 identified known ferroelectrics, 11 are reported to be plastic crystals. \cite{harada_2018,tang_multiaxial_2017, harada_directionally_2016,li_anomalously_2016,li_organic_2019,katrusiak_proton_2000,tang_organic_2020,
you_quinuclidinium_2017} Furthermore, the  high-temperature phases of BUJQIJ and BUJQOP have not been solved due to poor XRD data, \cite{ai_highest-tc_2020} which can indicate the high degree of disorder typical for plastic crystals. This shows that the screening procedure is suited to identify materials with plastic properties and orientationally disordered mesophases.

To identify plastic crystals amongst the candidate materials, we look to the CSD. We both investigated all structures within each refcode family and performed a structure search using Conquest\cite{bruno_new_2002} to find plastic phases reported with a different refcode. Three candidate materials were identified as plastic crystals where all molecules exhibit rotational disorder, namely TMAMBF11,\cite{mondal_metal-like_2020} ZZZVPE02,\cite{mondal_metal-like_2020} and ZISCUC.\cite{li_coexistence_2019}. LOLWEO has a reported high-temperature phase where one of the two molecular constituents shows rotational disorder. It can be noted that no reported plastic phase does not necessarily indicate that the material is not plastic -- only that no plastic phase has been studied. 

\subsection{Cohesive Energy and Thermal Stability}

For device applications, thermal stability is an important property, 
and operational temperatures should be significantly below melting and sublimation temperatures.
The melting temperature of plastic crystals can be higher than for similar ionic and molecular crystals.
As these materials transition into a plastic mesophase, the orientational disorder and increased freedom of movement increase the entropy,
making it less favorable to melt due to the reduced entropy gain,
as first reported by Timmermans \cite{timmermans_plastic_1961}.
The volume change at the phase transition is also small.\cite{staveley_1962}
In some cases, the reduced entropy gain can make sublimation more favorable than melting. 
A high Curie temperature is also a prerequisite for ferroelectric and piezoelectric applications. 

\begin{figure}
    \centering
    \includegraphics[scale=0.44]{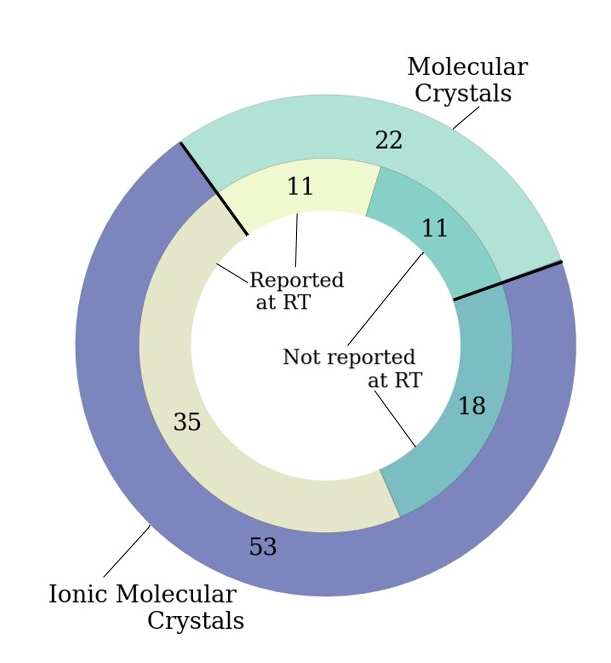}
    \caption{Overview of the molecular and ionic molecular crystals identified. The outer ring indicates the number of materials, and the inner ring indicates if the potentially ferroelectric phase is reported at room temperature.}
    \label{fig:RT_stable}
\end{figure}

For 44 out of the 75 materials identified in this study, the CSD entry concerns an ordered structure obtained at amibient temperature. The fraction of room temperature investigations is highest for ionic crystals, Fig.~\ref{fig:RT_stable}, a result that is in line with the higher stability of the ionic molecular crystals, due to electrostatic interactions.

\begin{figure}
    \centering
    \includegraphics[scale=0.25]{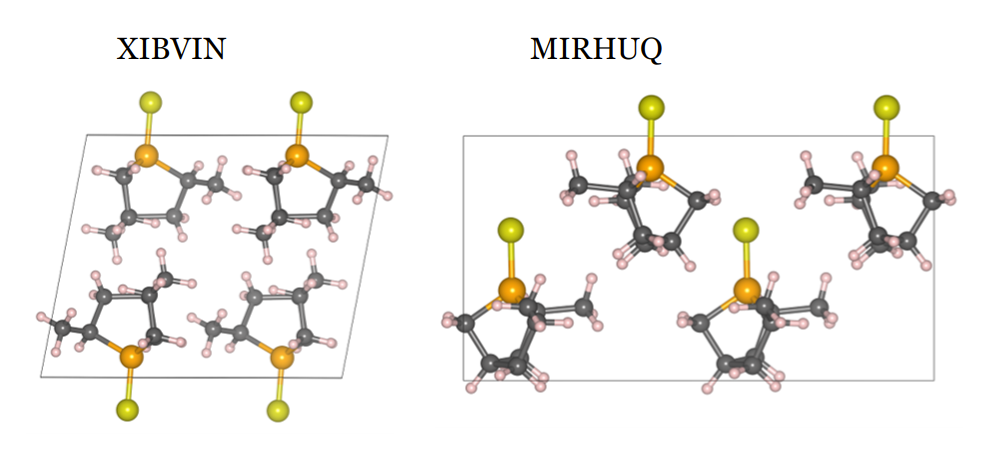}
    \caption{The two crystal structures of \ce{C8H15PS} identified in the screening. The alignment of dipoles differs in the two structures, with a low degree of alignment in XIBVIN, and better alignment in MIRHUQ.}
    \label{fig:align}
\end{figure}

To further investigate the cohesion of molecular crystals, we compute the cohesive energies of XIBVIN and MIRHUQ. These materials are polymorphs of the same compound, differing only by their alignment of molecules, see Fig.~\ref{fig:align}. 
XIBVIN has a low degree of alignment of $77^{\circ}$, combined with a low spontaneous polarization of $1.3~\mathrm{\mu C/cm^2}$. MIRHUQ shows better alignment, with angles of $29^{\circ}$ and $31^{\circ}$, and as one could expect, a higher spontaneous polarization of $11.3~\mathrm{\mu C/cm^2}$. While the larger polarization of MIRHUQ makes it more interesting as a ferroelectric, it is reported to sublimate at $388~$K, while XIBVIN melts at $443~$K. The computed values for the cohesive energies are $0.26~$eV/molecule for MIRHUQ and $0.30~$eV/molecule for XIBVIN, the larger value corresponding to the less aligned structure with the highest melting point. The difference in thermal stability also indicates that at most one of the structures has a plastic phase. If both structures transitioned into rotationally disordered plastic mesophase, it could be assumed that the plastic phases had similar structures, and thus similar phase transition temperatures and mechanisms. The higher phase transition temperature of XIBVIN indicates that this structure transitions into a plastic mesophase, as the entropy gain in such transition can stabilize the solid phase.

\subsection{Electronic Band Gap and Multisource Energy Harvesting}

Ferroelectric materials with band gaps in the visible range can be of particular interest, as these can be promising for application in multi-source energy harvesting devices where piezo- or pyroelectric energy harvesting is combined with the harvesting of solar energy. \cite{li_semiconducting_2017,bai_energy_2018} Therefore, we computed the band gaps for all identified materials to find those
that could be suitable for such applications.

Tables \ref{known_results} and \ref{candidate_results1} lists the computed band gaps. As the computations were performed using the vdW-DF-cx functional, 
which describes exchange at the generalized-gradient approximation level, \cite{berland_exchange_2014}
we expect the values to be underestimated. Three materials are predicted to not have a band gap, this was  confirmed using the hybrid functional HSE06.
Most of the computed band gaps exceed $\mathrm{4~eV}$;  
however, six materials have values smaller than $\mathrm{2~eV}$.
Four of these belong to Trimethyl-X-Y group, which would be interesting for multi-source energy harvesting. 

\subsection{Design Strategies for Ferroelectric Plastic Crystals}
Novel and improved ferroelectric plastic crystals can be engineered by making new combinations of molecular species. 
As such, the various species found in the various plastic crystal candidates can be used as building blocks for novel material design, a selection of these are displayed in Fig.~\ref{fig:mols}.

Of particular interest are substitutions that are likely to result in similar or  
isostructural materials. 
For systems where this is possible, solid-solution engineering can be used to tweak material properties. Three examples of pairs of candidates for such alterations are found among the trimethyl-X-Y materials. The compositions between each pair of materials are similar, only differing by the substitution of molecules with similar geometries. 
PEVXOE and PEVXUK are isostructural and only differ by the substitution of \ce{(CH3)4P+} for  \ce{(CH3)4As+}.
For ZZZVPE02 and TMAMBF11, the substitution of \ce{(CH3)3BH3N} to \ce{(CH3)3BF3N} yields an isostructural material where the spontaneous polarization is increased from $5.4$ to $20.3~\mathrm{\mu C/cm^2}$. A similar effect is seen for YODGON and VUGNUG, where the substitution of \ce{OCN-} for \ce{N3-} leads to an increase in the spontaneous polarization from $6.4$ to $22.0~\mathrm{\mu C/cm^2}$. 

Another example of substitution resulting in similar packing is the earlier reported ferroelectrics TEDAPC28 and SIWKEP, which consist of a dabco molecule and \ce{ClO4-} or \ce{ReO4-}, respectively. Such a substitution somewhat changes the
orientation of molecules, resulting in a different space group. 
Whereas the \ce{ReO4-} material has a computed spontaneous polarization of $8~\mathrm{\mu C/cm^2}$, while the \ce{ClO4-} analog has a value of $5.9~\mathrm{\mu C/cm^2}$. 
The screening identified ten different tetrahedral inorganic anions, but other options also exist and can be considered for the design of new ferroelectric plastic crystals. 

Substitutions of single atoms in the organic globular molecule can also be a route to engineer the ferroelectric properties.\cite{liu_molecular_2020} 
For instance, Lin et al. substituted a hydrogen atom in tetramethylammonium for the halogens I, Cl, and Br. The crystal structure was retained, while the number of ferroelectric polar axes increased from 2 for iodine to 6 for chlorine.\cite{lin_halogen_2023}

\section{Conclusions}

Using a CSD-based workflow, we have identified 55 candidate ferroelectric plastic crystals. 
The 21 with spontaneous polarization exceeding $\mathrm{10~\mu C/ cm^2}$ are arguably the ones with the most potential in ferroelectric devices.
Among these, eight in the trimethyl-X-Y group also display a large variation in electronic band gaps that could make them useful also for multi-source energy harvesting.

Our study has successfully identified a range of candidate ferroelectric plastic crystals, including 16  that have been reported as ferroelectrics in the past.
The screening criteria used were quite strict, and relaxing some of them, such as the size limitations on both molecules and unit cells, would have expanded the pool. 
The criteria were based on the previously reported ferroelectric plastic crystals. For this reason, crystal structures that differ significantly from the earlier reported ones could have been overlooked.
Still, the identification of the boron cluster-based materials was unexpected, illustrating the potency of the CSD screening. 
In the design of novel ferroelectric plastic crystals, a good starting point is combining different globular molecular species, i.e., combinations of cations and anions. The various molecular species in the materials identified here, and the corresponding crystal structures, can serve as inspiration for such design.  

In this study, we have not assessed whether the polarizations of the candidate ferroelectrics are switchable, which is a criterion for ferroelectricity. 
However, all the candidate materials will have pyro- and piezoelectric properties due to their crystal symmetries. 
This study and the identified compounds should stimulate further theoretical or experimental studies, to assess their ferroelectric switchability, and other characteristics such as the Curie temperatures, and material stability. 

\section{Data availability}
All the relaxed crystal structures used in this study can be accessed through the NOMAD database at \url{https://dx.doi.org/10.17172/NOMAD/2023.06.12-1}. All other data is available upon reasonable request.

\begin{acknowledgments}
The computations of this work were carried out on UNINETT Sigma2 high-performance computing resources (grant NN9650K). This work is supported by the Research Council of Norway as a part of the Young Research Talent project FOX (302362). Structure and molecule figures are made using the software programs Mercury and VESTA. 
\end{acknowledgments}

\appendix
\section{\label{sec:appendix}Molecular Graph Theory}
\begin{figure}[ht]
    \centering
    \includegraphics[scale=0.25]{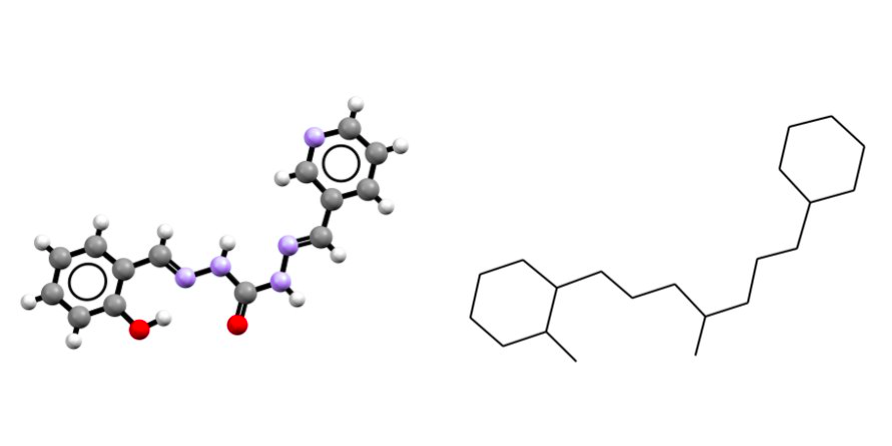}
    \caption{Illustration of a network representation of a molecule. The network only considers the connectivity, as given by covalent bonds between atoms. Hydrogen atoms are omitted when constructing the molecular graph.}
    \label{fig:network}
\end{figure}  
A graph, or network, consists of a set of nodes that can be connected by edges. \cite{iniguez_bridging_2020}
Fig.~\ref{fig:network} illustrates this in the context of a molecule. A covalently bonded molecule is represented as a graph, with atoms as nodes and covalent bonds as edges. 
A node with two edges is called a chain and if a part of a graph can become disconnected by cutting one single chain, it is called a bridge. 
To avoid long, flexible molecules that are unlikely to rotate, we required molecules to have at most two connected bridge nodes, i.e., the molecule shown in Fig.~\ref{fig:network} was excluded due to a high number of connected bridges.

\bibliography{references}

\end{document}